# Deployment of Microcells by integrating LTE-U with LTE

**Deven Panchal**
*Student, Dept. Of Electrical and Computer Engineering,*
*Georgia Institute Of Technology, Atlanta, USA.*
devenrpanchal@gatech.edu

**ABSTRACT**

With increasing data requirements of users, cellular operators are finding new ways to fulfil these requirements. These attempts involve the practice of deploying Wi-Fi access points nearer to the user and backhauling it to the nearest eNB (in case of LTE and LTE-A).

The paper studies LTE-U, an extension of LTE which works in the unlicensed spectrum, as a potential solution to this problem. It is based on the idea of densification. Network deployments incorporating LTE-U will be able to better cater to the growing data rate demand of voice and video, thus reducing the load on eNB.

Further we explore the possibility of LTE-U as an alternative to Wi-Fi or co-existing with Wi-Fi deployments and issues revolving around this idea. We show that LTE-U deployment solves the problem of capacity in both cases.

## I. INTRODUCTION

Increased capabilities of handheld devices have led to emergence of high data rate applications like video streaming, online gaming etc. that can be run on them. To support these applications, the network must be able to offer increased capacity at a reduced latency. However, this is not always possible because of spectral limitations. Not only is finding unused licensed spectrum itself difficult, but the cost of this spectrum is also very high. So operators usually strike out this option as a solution to increase capacity.

Although LTE-U has many advantages, the idea of LTE-U is based upon two fundamental problems. The problem of scarcity of spectrum and the problem of ever-growing data rates.

Section II talks about why we need a technology like LTE-U. Section III talks about the basics of LTE-U technology. Section IV briefly talks about the implementation of LTE-U and its architecture. Section V talks about handoffs in LTE-U and the simulation of a fuzzy logic based handoff decision algorithm. Section VI talks about the co-existence of LTE-U and Wi-Fi. Section VII briefly talks about authorization in LTE-U.

## II. THE PROMISE OF LTE-U

The amount of unlicensed spectrum assigned or currently planned to be assigned is comparable to or even more than the amount of licensed spectrum. This is shown in Fig 1. One idea to utilize this spectrum is to offload some cellular traffic to Wi-Fi APs. However, such efforts have not been achieving the expected targets in terms of network performance improvement or cost reduction for operators due to several reasons.

Thus LTE-Unlicensed (LTE-U) becomes a very strong candidate to solve this problems due to the following reasons.

1. LTE is currently the most advanced mobile telecommunication technology. Many operators all over the world are upgrading their networks to LTE and LTE-A. But the problem of coverage and latency still exists. Hence there is a need of a technology which can extend the LTE coverage and yet seamlessly integrate with the LTE system ultimately giving the best performance. This is what LTE-U does.

2. For the operator having a unified network for both licensed and unlicensed spectrum and being able to better integrate the unlicensed access to improve system capacity as a whole is an advantage. LTE-U deployment would be as simple as deploying

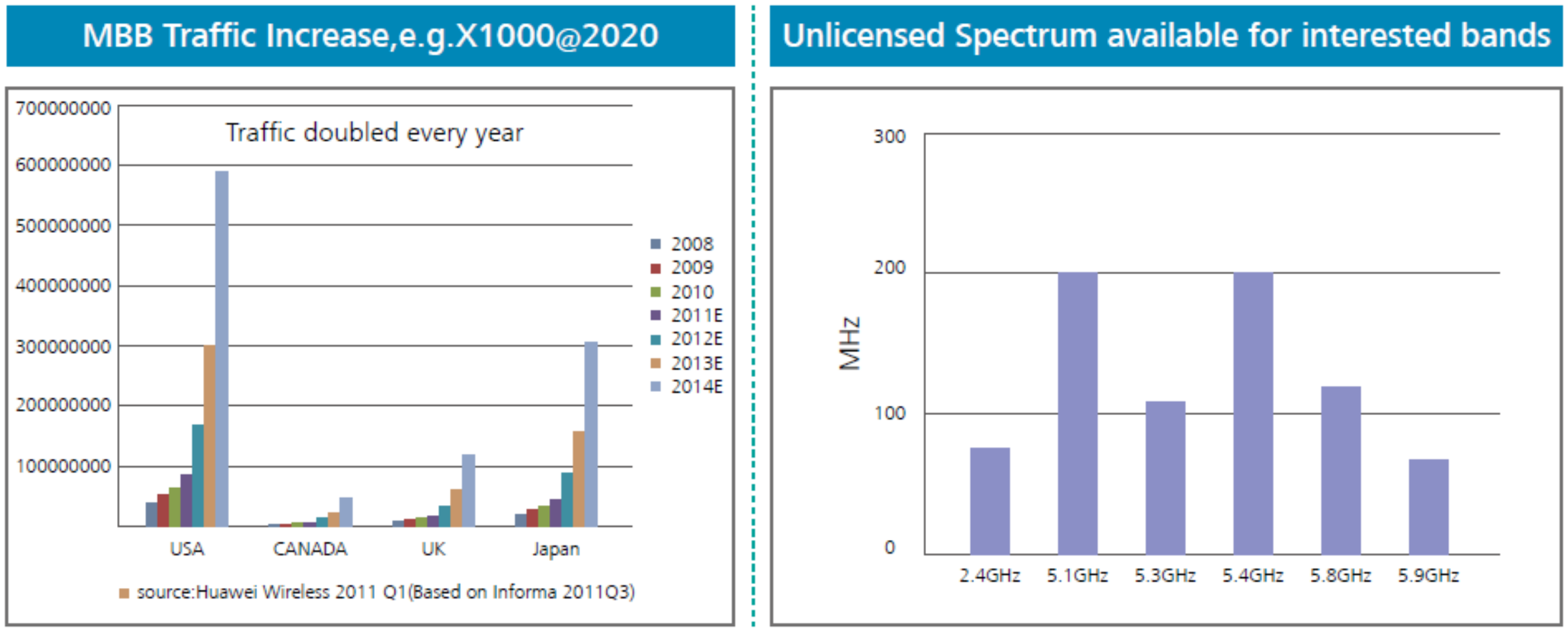

**Fig 1: Traffic increase and available unlicensed spectrum [3]**

LTE-U wireless access points and performing a software upgrade in the eNB because both LTE and LTE-U would have:

a) Common RAN framework
b) Common core network
c) Common operations and management systems,
d) Common charging, authentication, acquisition, access, registration, paging and mobility procedures [3].

Thus deployment and management of the system as whole becomes easy and profitable.

3. For the user, good coverage due to dense deployment and good QoS due to carrier aggregation whenever unlicensed frequency is available. Also there is a guarantee of service due to anchor in the licensed spectrum, where all the control and signaling is carried out in the licensed spectrum as we will see when we understand the LTE-U technology.

## III. INTRODUCTION TO LTE-U TECHNOLOGY

In LTE, support for wider bandwidth is provided through aggregation of multiple component carriers [6]. The networks and devices co-ordinate for opportunistic access and aggregation of unlicensed spectrum without manual intervention. Users are able to transmit and receive on several component carriers simultaneously. For example, 4×20 MHz carriers may be aggregated to provide 80 MHz of spectrum to a user in the downlink. Currently, up to five carriers may be aggregated to provide users up to 100 MHz of bandwidth. Carrier aggregation configuration is user-specific. For example, a system may have four 20-MHz downlink carriers and two 20-MHz uplink carriers available. Individual users, however, can be configured to support different subsets of this system configuration. For example, one user may be configured to support DL: 2×20-MHz and UL: 1×20-MHz while another user may be configured to support DL: 4×20-MHz and UL: 2×20-MHz. This user-specific configuration is up to system implementation and user capability. LTE allows fast activation and deactivation of carriers so that this user specific configuration can be dynamically managed. Each user will be configured with one downlink and one uplink Primary Component Carriers (PCCs). The remaining carriers are called Secondary Component Carriers (SCCs). Each component carrier will handle a separate data stream that is aggregated or segmented at the MAC. Separate HARQ processing and associated control signaling is required for each of the component carrier. This allows separate link adaptation and MIMO support for each carrier, which should improve throughput since each data transmission can be independently matched to channel conditions on each carrier. Cross-carrier scheduling is supported, i.e. scheduling data transmission from another carrier. Uplink control signaling, however, must be transmitted on the uplink PCC only. The Carrier Aggregation (CA) mechanisms can perform the flexible offload between licensed and unlicensed carriers, where the unlicensed carriers are operated as Secondary Carriers associated to and controlled by the existing licensed LTE Primary Carriers. Regarding the possible operating modes, the unlicensed spectrum can be operated as a TDD (DL+UL) carrier or a DL-only carrier, regardless of the operating mode of primary component carrier in the licensed spectrum.

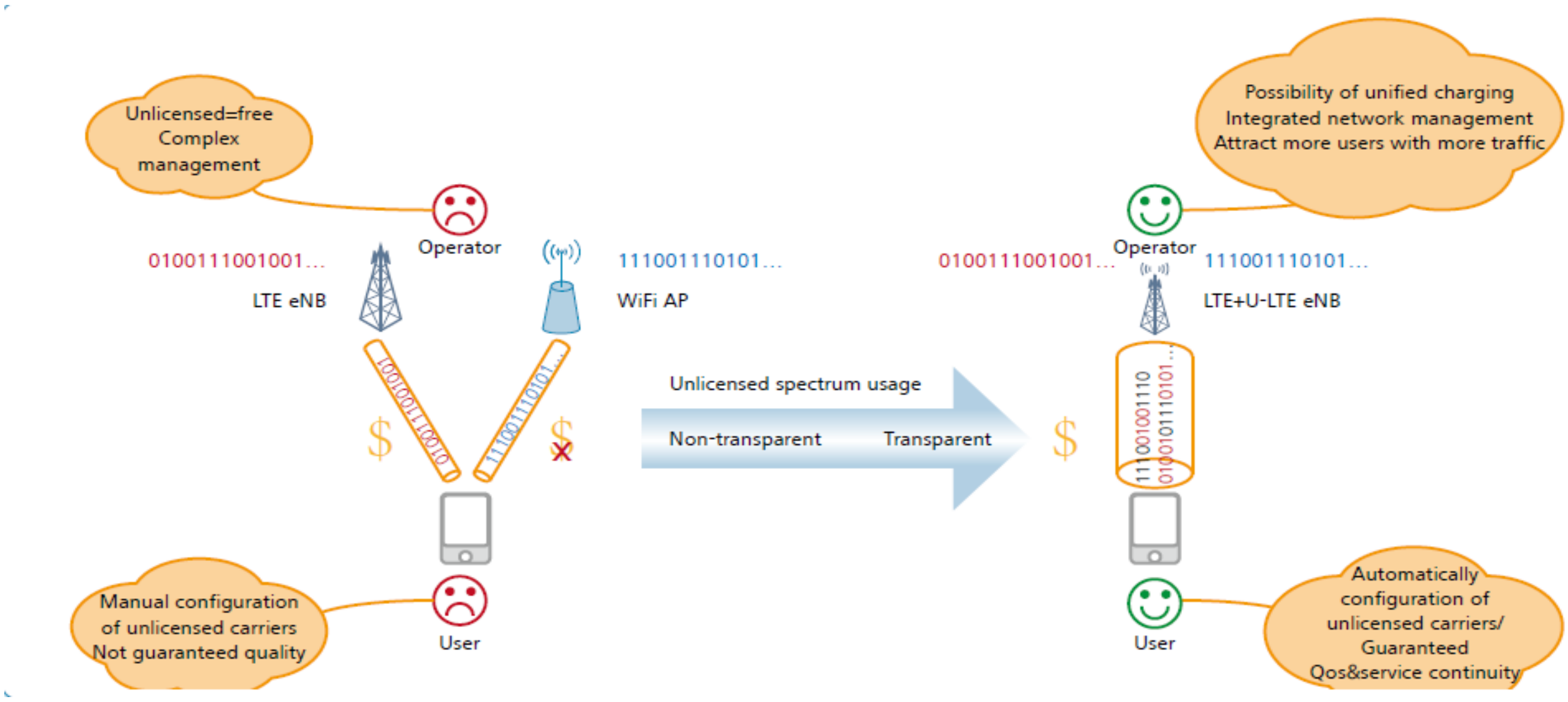

**Fig 2: Handling traffic volume by the network.**

The main concept of LTE-U is to aggregate carriers in licensed and unlicensed bands. A user is configured with a PCC in the licensed band and several SCCs in the unlicensed band. All control-plane and L1 control signaling (e.g. ACK/NACK, CQI) are carried on the primary component carrier. User-plane data can be transmitted either on the primary or secondary component carriers [6]. As an integrated part of the LTE system, the enabling or disabling of the unlicensed secondary carriers can be seamlessly controlled by the network without need of manual configuration by the user. In this way, LTE-U can enable traffic volume to be carried on an unlicensed or licensed carrier in a transparent way from the user perspective as shown in Fig 2.

Since the 2.4GHz band is already crowded with residential and even public deployments, the 5GHz band is the main candidate in terms of relatively large amounts of unlicensed spectrum with globally common availability, as well as relatively good channel propagation performance. Within the 5GHz band, the 5150-5250MHz and 5250-5350MHz blocks are also widely used by residential WLAN in indoor/outdoor scenarios and hence these blocks are not preferred. The candidate band should allow for both indoor and outdoor deployment and transmit power should be high enough as operator deployment is the main target scenarios for U-LTE.

Since LTE-U would be operated as a Secondary cell controlled by a aggregated licensed Primary cell, the band combination should be chosen to avoid strong intermodulation interference between each other, especially in-device interference for the given unlicensed band (assuming the licensed component band need not be selected to represent the most likely deployment cases but rather the cases that are most common across different regions). For example, though 5G Hz band is supposed to be well isolated from current IMT cellular bands, some cross-band emission issues (e.g., the inter-modulation interference between 5470~5725MHz band and the 1.8GHz band) may prohibit their aggregation.

| Sub-bands | 5150-5250MHz | 5250-5350MHz | 5470-5725MHz | 5725-5825MHz |
|---|---|---|---|---|
| EIRP | 17dBm/23dBm | 23dBm/30dBm | 23dBm/30dBm | 23dBm/30dBm/36dBm |
| US/Canada | Indoor | Indoor/Outoor | Indoor/Outoor*3 | Indoor/Outoor*1 |
| EU | Indoor | Indoor | Indoor/Outoor | NA |
| Korea | Indoor | Indoor/Outoor | Indoor/Outoor*4 | Indoor/Outoor |
| Japan | Indoor | Indoor | Indoor/Outoor | NA |
| China | Indoor | Indoor | NA | Indoor/Outoor*1,*2 |
| Australia | Indoor | Indoor/Outoor | Indoor/Outoor | Indoor/Outoor |
| India | Indoor | Indoor | NA | Indoor/Outoor |
| Inter-modulation interf with licensed bands | 2.6GHz,1.7GHz 800MHz | 2.6GHz,1.7GHz 800MHz | 1.8GHz,900MHz | 1.9GHz,1.4GHz |

**Table 1: 5 GHz sub-band deployment permission and band combination analysis [3]**

## IV. LTE-U ARCHITECTURE

As per [3], the following changes will need to be made to the LTE Air Interface when moving to LTE-U. They are as follows:

1. Enhancements of CA mechanism to facilitate opportunistic use of channel
2. Adapting LTE frame structure with respect to LBT requirements
3. Support of UL transmission
4. Enhanced interference coordination with respect to more complex interference scenarios in unlicensed spectrum

LTE-U will need upgrades in the eNB but most of the network entities will remain common between LTE and LTE-U as mentioned in section II. Also as explained earlier, this leads to lower costs for operators.

In this section, we review the resulting architecture of LTE-U. A new resource handler mechanism, **LTE-U Gateway** which acts as an intermediary between macrocell and a group of LTE-U cells and takes care of the handoff between them. In Fig 3, HeNB-GW represents LTE-U Gateway. It also provides the functionality of Mobile Management Entity (MME) to LTE-U microcells.

The LTE-U GW appears as macrocell to the MME. It appears to the LTE-U cells as a MME. The S1 interface between the LTE-U cell and the core network is the same, regardless whether the LTE-U cell is connected to the core network through a LTE-U GW or not. LTE-U GW has to connect to the core network for serving UEs mobility between cells, which does not require inter MME handovers.

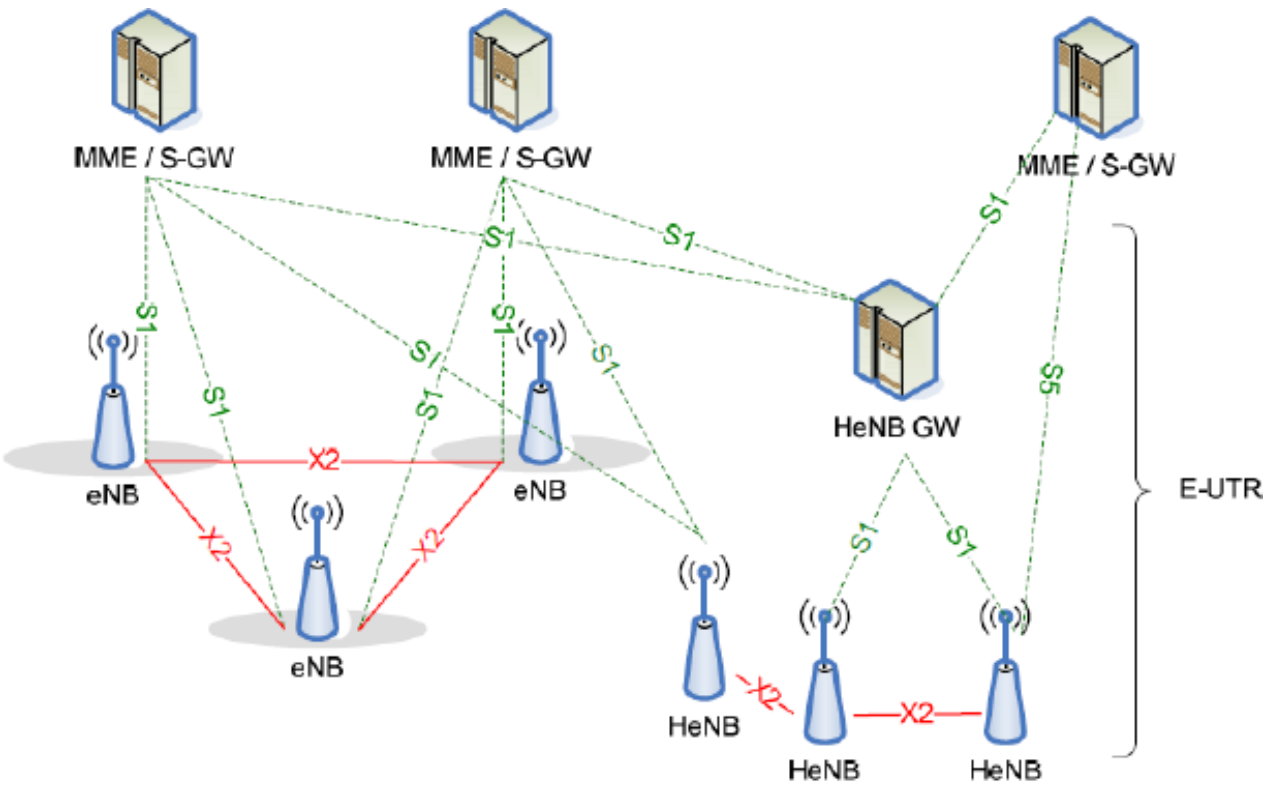

**Fig 3: Overall E-UTRAN architecture with LTE-U Gateway (HeNB GW)**

## V. HANDOFF IN LTE-U

The figure below shows typical handoff procedure in an LTE system.

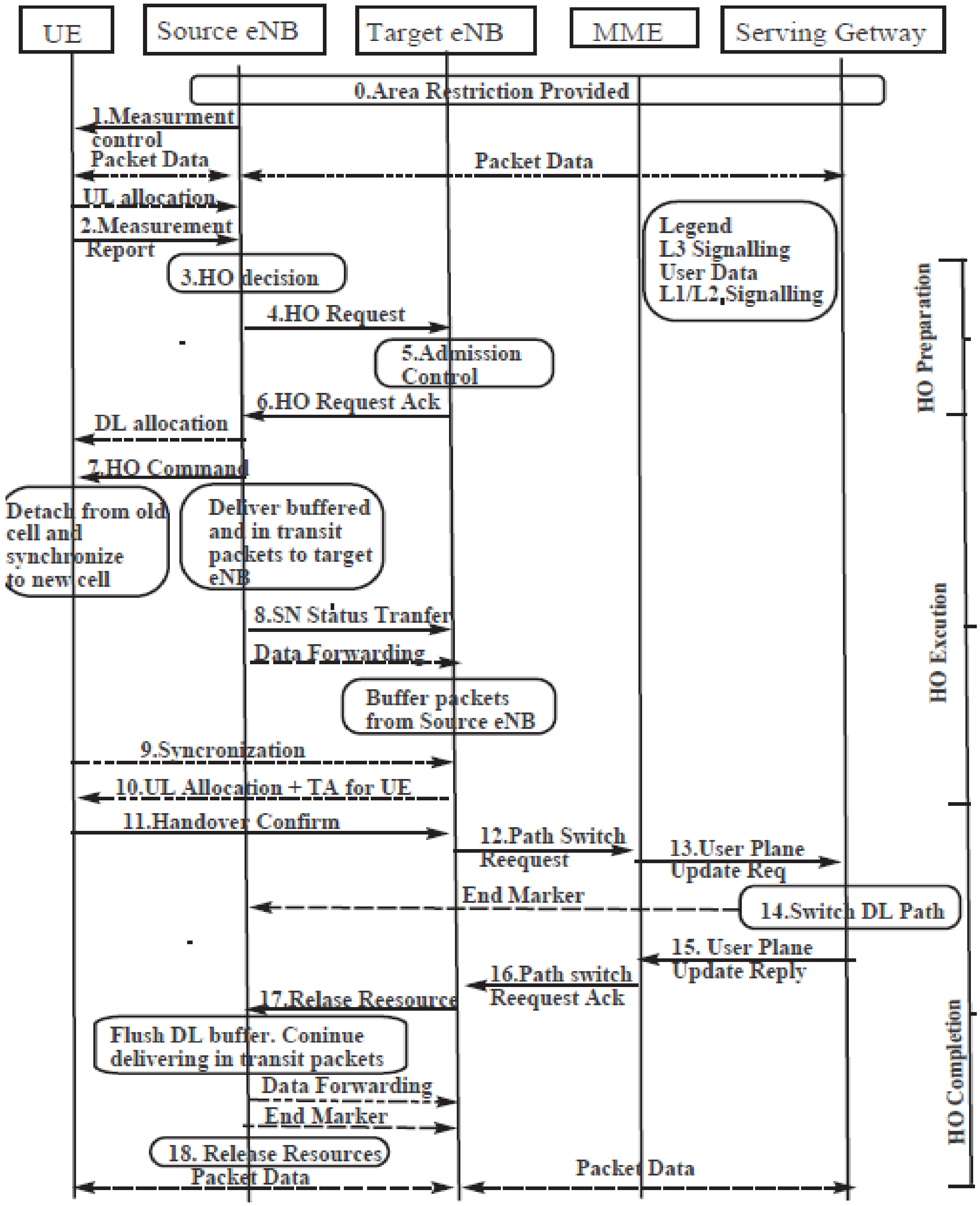

**Fig 4: Typical LTE Handover Procedure [4]**

*Simulation of the Handoff Decision algorithm:* A Fuzzy logic based approach was chosen to make the Handoff decision because fuzzy logic can deal with multiple imprecise yet correlated input data (which is the case in a typical cellular environment) to make intelligent decisions. A fuzzy logic system consists of 3 main parts:

1. *Fuzzifier:* Assigns degrees of membership in different classes to the different inputs.
2. *Fuzzy Inference System (Mamdani):* It is the 'brain' which operates according to specified fuzzy rules.
3. *Defuzzifier:* Combines fuzzy outputs into discrete values that can be supplied to control.

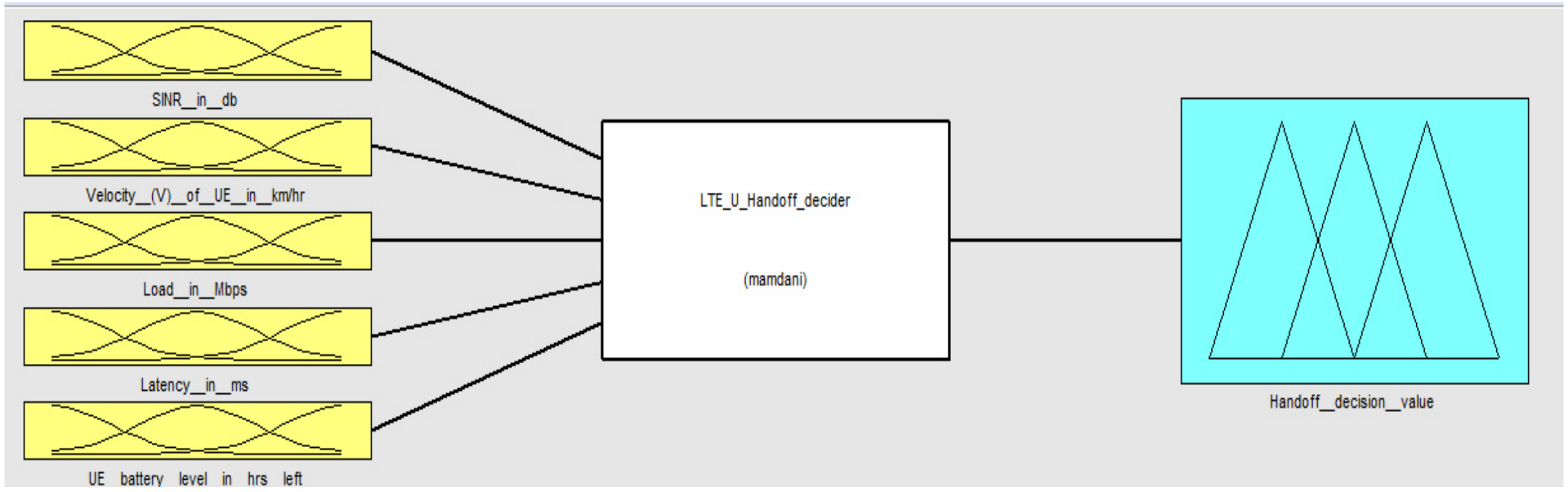

**Fig 5: Fuzzy logic system implemented in MATLAB**

*Inputs:* Our handoff decision algorithm takes into account the following parameters to decide whether or not to perform handoff.

1. *SINR:* SINR is chosen as a measurement of UE as handover decision criteria as it gives the channel quality information. If channel quality is good, UE will have better quality of service for real time traffic (voice).

2. *Velocity of UE:* In order to prevent unnecessary handovers, the velocity with which UE is travelling is also taken into account. If we know the UE movement, towards where it is heading, we can decide whether to take proactive or reactive approach. The mobility prediction uses UE's physical movement and estimates its next position. If we know the next position, the system can decide how to make a handover. The unnecessary handovers are mitigated by using reactive handover mechanism, where the handover is postponed until the UE reaches to a threshold signal strength of the target cell.

3. *Authorization Access:* Discussed in Section V, it validates the handover only for those UE who have access to the LTE-U microcell. Although a UE can request for a temporary access to the microcell also. For the purpose of simulation, it was assumed that all the users have authorization access.

4. *Latency:* Latency requirement defines the type of the application being run on the UE. While voice, video, online gaming require low latency, there is no such requirement on data. The latency required or the type of application being run on the UE has been used as an input to the handoff decision system. In a scenario where the SINR of another eNB or HeNB is comparable to that of the current eNB or HeNB, and there is a possibility of a low latency link from the new eNB or HeNB, it may make sense to handoff.

5. *UE battery level:* The UE also reports its battery level to the native eNB so that the eNB can decide whether or not UE would be able to sustain a handoff and whether it worth handing off considering the battery level.

6. *Load at the new cell:* This parameter is used to check whether the new cell has enough resources to handle the user for which the handoff has been performed.

*Decision:* Decision of the fuzzy inference engine is based on some fuzzy rule base. One of the rules we had defined is as shown below in the verbose representation.

```
If (SINR__in__db is near_RTH) or
(Velocity__(V)__of__UE__in__km/hr is
high) or (Load__in__Mbps is low) or
(Latency__in__ms is voice/gaming) or
(UE__battery__level__in__hrs__left is
high) then (Handoff__decision__value is
handoff) (1)
```

*Output and its interpretation:* The output of the simulation is a number which would help decide whether to handoff or not. The simulation was set up such that if the resultant output number is nearer to 1 it means that the fuzzy logic based decider is recommending a handoff. If the resultant output number is nearer to 0, we don’t need to handoff. The scenario in Fig.6 shows requires a handoff as is indicated by the Handoff decision value=0.853.

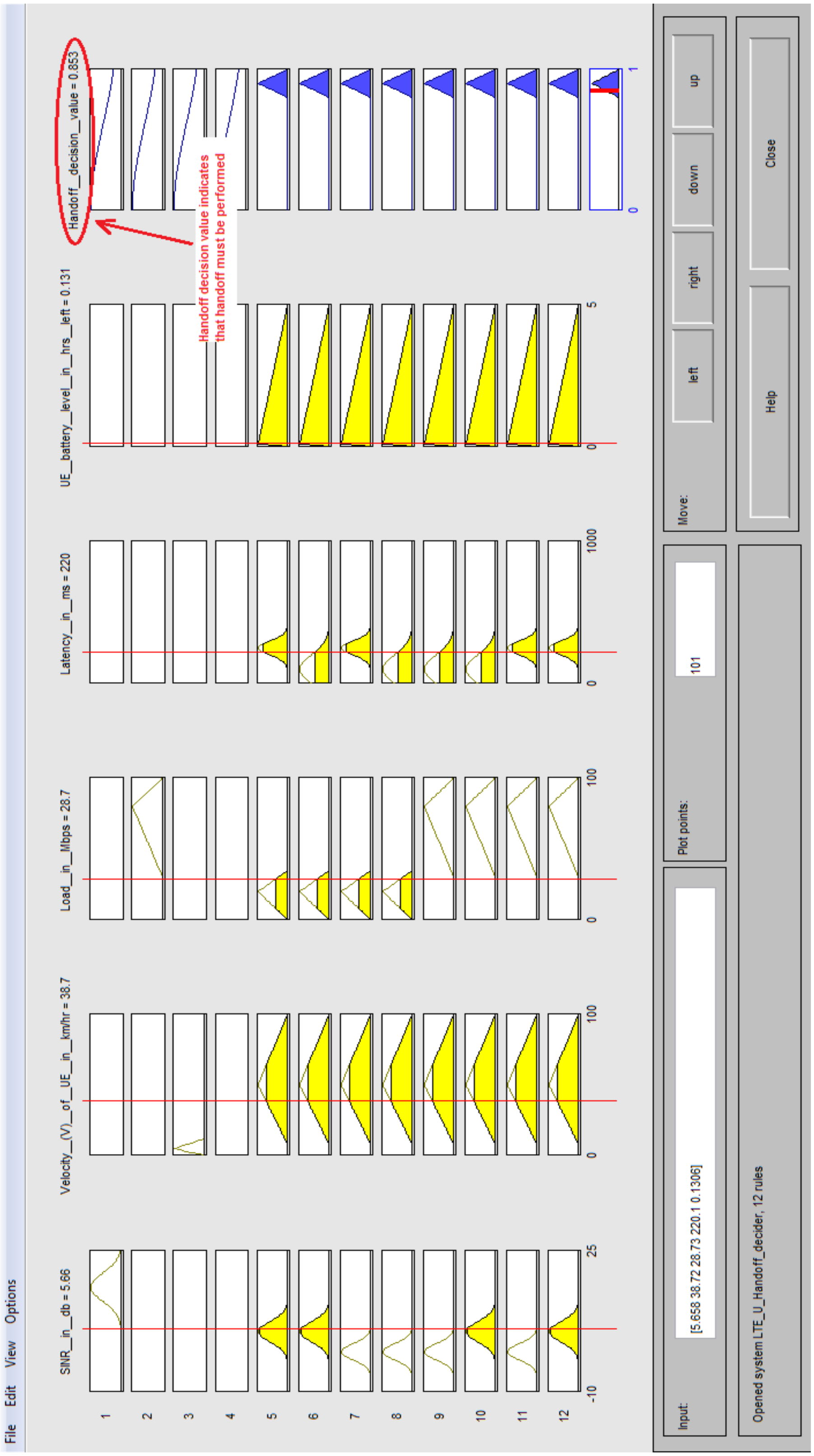

**Fig 6: Results of the simulation of fuzzy logic based simulation algorithm for handover decisions. Notice how for particular scenario handoff values is >0.8 indicating handoff is necessary.**

A handoff in an LTE-U system does not always mean a HeNB-HeNB handoff. It may mean one of the following three handover scenarios:

1. *Macro-Macro Handover:* This scenario is triggered when the SINR of the target eNB is better than that of the serving eNB plus the HHM. When UE is moving from one macrocell region to another macrocell region with more speed (velocity greater than 10kmph), target macrocell is predicted based on the Euclidean distance between UE and neighbor macrocells. After deciding target cell, we need to check for the type of application the UE is using. If UE is using real time traffic (voice), then perform proactive handover mechanism. If UE is using non-real traffic then perform reactive handover mechanism. In case the UE is moving with a velocity less than 10kmph then no mobility prediction is used. The type of handover is decided based on the type of application, i.e. proactive handover for real time traffic or reactive handover for non-real time traffic.

2. *Macro-LTE-U Handover:* This scenario is triggered when the SINR of the target HeNB is better than that of the serving eNB plus the HHM. If UE is moving with more speed (velocity greater than 10kmph), then don't perform the handover from macrocell to femtocell, as UE may quickly move out of femtocell coverage region, which may lead to an unnecessary handover. If the UE moves with less speed (velocity lesss than 10kmph) then perform the handover, as UE is expected to stay for some period within the femtocell coverage region. The handover performed is based on the type of application as in the previous case.

3. *LTE-U – Macro Handover:* In this scenario, handover is performed when UE experiences signal strength from macrocell which is greater than signal strength of the femtocell by a margin of HHM. The exact type of handover is again based on the type of application.

Although in practice we recommend using the fuzzy logic-based approach discussed earlier for all the above type of handovers, for the purpose of clarity to the reader and to avoid the inconvenience of representing the many fuzzy rules, we present a straightforward algorithm which can be used to make handover decisions in above cases. The parameters for simulation may be chosen as suggested in Table 1 or they may be changed to suit specific interests.

| **Parameters** | **Values** |
|---|---|
| Cellular Layout | Hexagonal grid, 7 clusters/cell |
| Cell Radius | 2000 m |
| Propagation Model | Okumura Hata Model |
| Traffic Load | Voice and Data |
| Path Loss(in macrocell) | 15.3+37.6log10(R[m]) |
| Log Normal Shadowing(macrocell) | 8 dB |
| BS TX Power(macrocell) | 43 dBm |
| BS TX Power(microcell) | 10 dBm |
| UE velocity | Random |
| Number of macrocells | 4 |
| Number of microcells | 15/cell |
| Number of UEs | 20 |
| Simulation Time | 100 seconds |
| Application Type(Voice or data) | Random |
| Authorization Access | Given to all users |

**Table 2: Simulation Parameters [7]**

*Algorithm:*

```
Input: SINR value of serving cell
Input: V speed of the UE
Input: HHM Handover Hysteresis Margin

if (SINR(Target-eNB) > SINR(Serving-
eNB) + HHM) then
      if (V > 10) then
          predict_user_mobility;
               if (traffic_used =
real_time) then

proactive_handoff;
              else
                      reactive_handoff;
              end
       else
              if (traffic_used =
real_time) then
                  proactive_handoff;
             else
                  reactive_handoff;
             end
      end
end

if (SINR(Target-LTE-U Gateway) >
SINR(Serving-eNB) + HHM) then
      if (V > 10) then
            no_handoff;
      else
             if(Auth_access =1)
                   if (traffic_used=
```

```
real_time) then

proactive_handoff;
                  else

reactive_handoff;
                  end
            else
                 request_temp_access;
            end
end
if (SINR(Target-LTE-U cell) >
SINR(Serving-LTE-U cell) + HHM) then
        if (Auth_access =1)
                     if (traffic_used =
real_time) then

proactive_handoff;
                    else

reactive_handoff;
                    end
           else
                request_temp_access;
           end
end

if (SINR(Target-eNB) > SINR(Serving-
LTE-U Gateway) + HHM) then
           if (traffic_used =real_time)
then
                    proactive_handoff;
          else
                    reactive_handoff;
          end
end
```

## VI. CO-EXISTENCE OF LTE-U AND WI-FI AND RELATED ISSUES

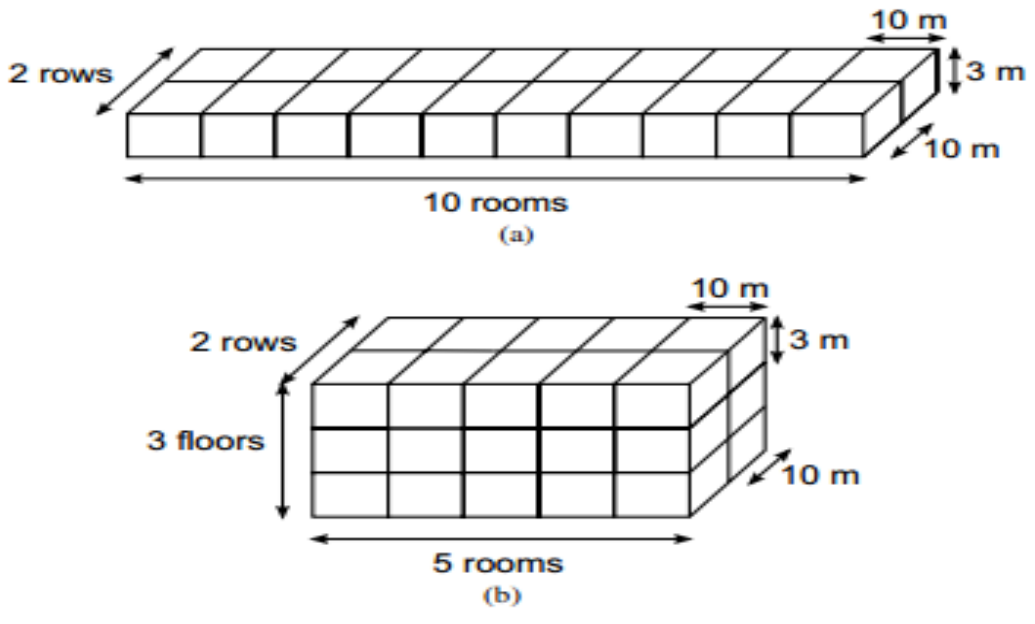

**Fig 7: (a) Single Floor Scenario, (b) Multi-floor scenario**

Having gone through the LTE-U technology in detail, we can better appreciate the real advantages of LTE-U when compared with Wi-Fi.

Firstly, LTE and Wi-Fi are different technologies, and so their interworking is complex.

Secondly, LTE-U which is based upon LTE. LTE is a more evolved technology with a better air interface with provisions for better coping with interference and also providing seamless mobility. LTE itself has many technologies as a part of it, which make it favorable as a co-existence technology. For eg. Listen-Before-Talk (LBT) scheme is used to deal with interference from IMT systems. It involves occupying the channel only after quickly sensing it. The Dynamic Frequency Selection (DFS) technology in LTE takes care of interference with non-IMT systems. Transmission Power Control (TPC) is used to limit the power from leaking into the neighboring bands and affecting other deployments.

There are many other reasons including the fact that LTE has a more robust control channel protection compared to Wi-Fi. Other LTE features like closed loop-link adaptation (as opposed to open-loop link adaptation in Wi-Fi) and HARQ retransmission scheme make the LTE PHY and MAC layers more robust.

But since LTE-U and Wi-Fi will have to co-exist considering large scale Wi-Fi deployments, the issue of interference between the two systems is important.

A study on the performance of co-existence between Wi-Fi and LTE-U for indoor deployment scenarios was conducted by Nokia Research Center, Berkeley, USA [11]. Fig 7 shows the two deployment scenarios. It has been shown in various studies by Huawei that similar results are obtained in case of outdoor deployments. Results of the study where the user throughput and the SINR of both the technologies for both the single floor and multi-floor scenarios -

*For single floor scenario:*

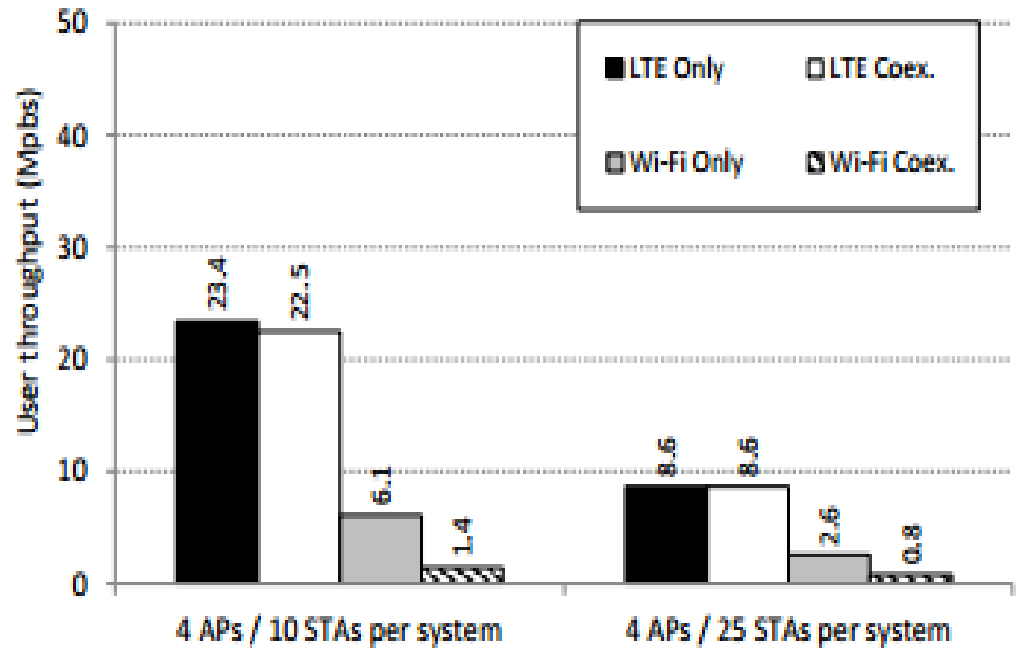

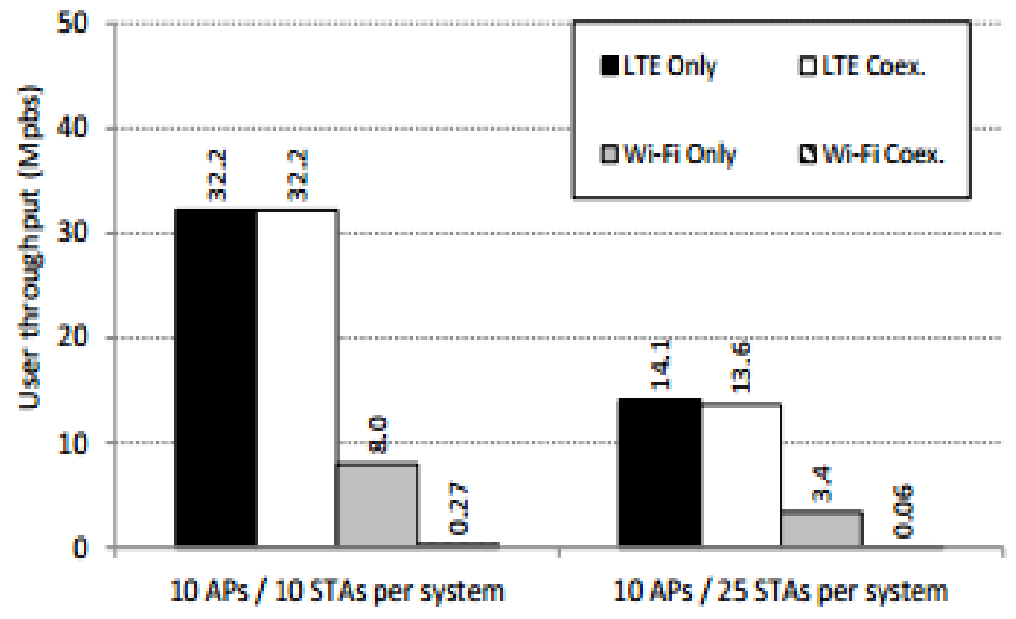

**Fig 8: (a) Sparse Deployment (b) Dense Deployment**

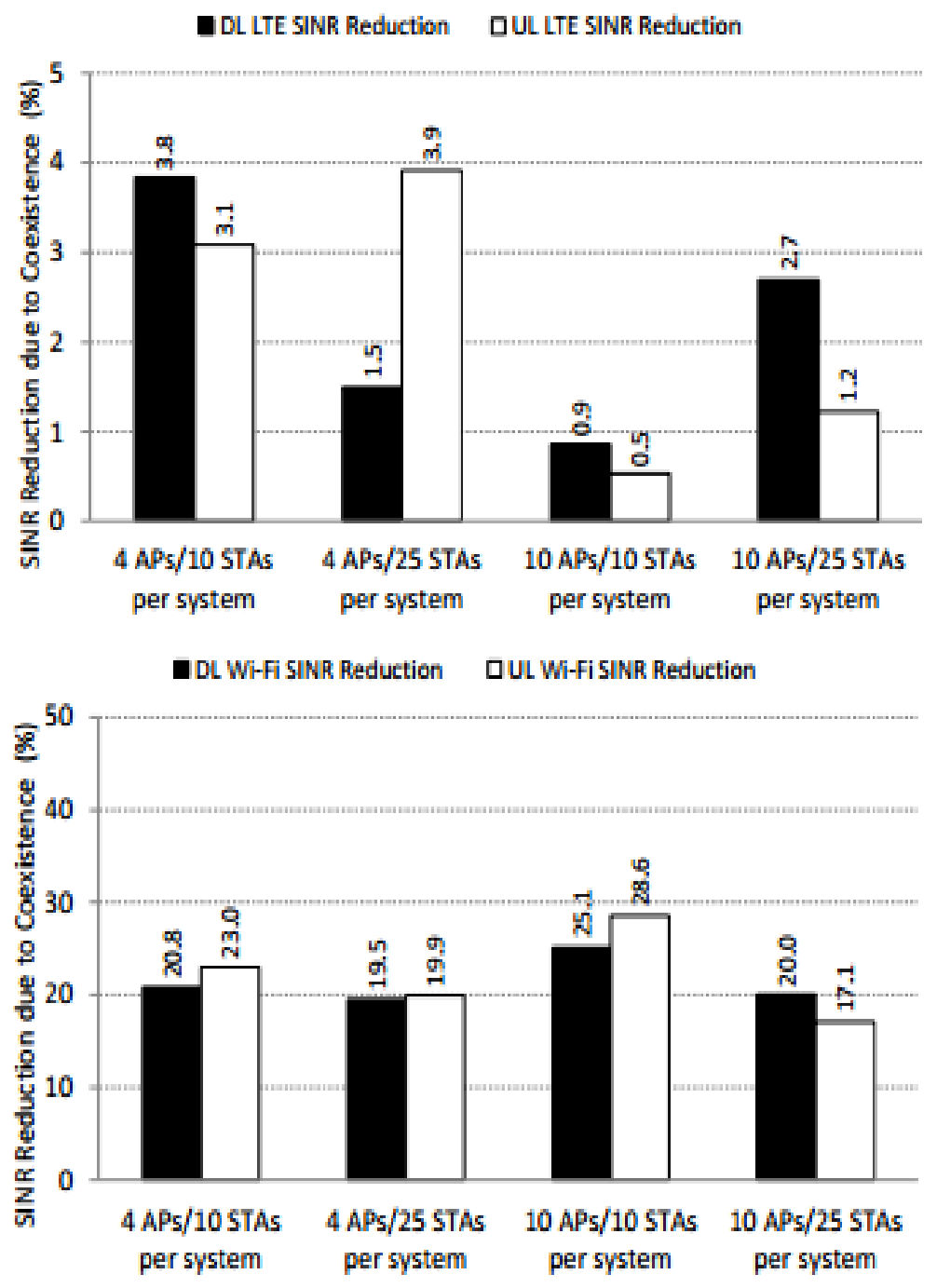

**Fig 9: (a) LTE SINR performance (b) Wi-Fi SINR Performance**

*For Multi Floor Scenario:*

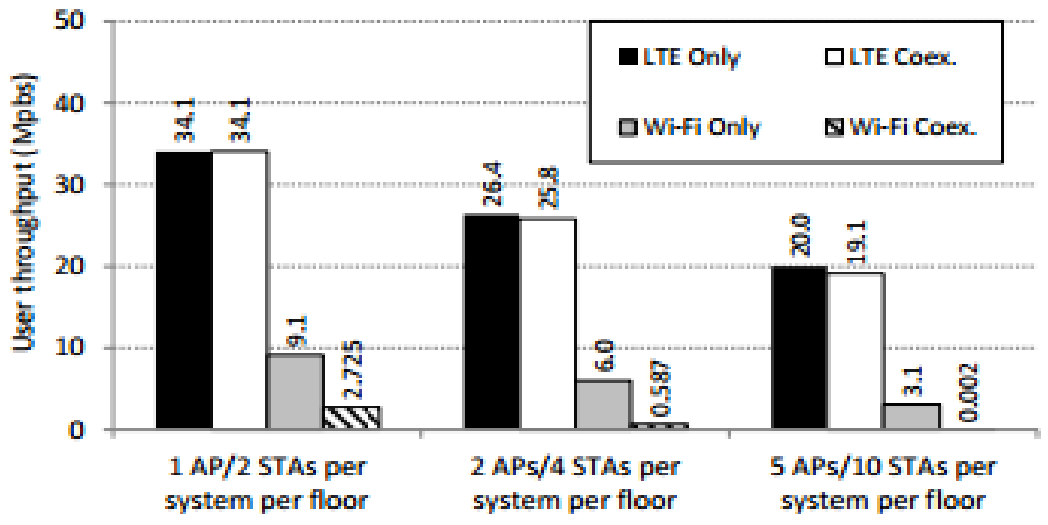

**Fig 10: Throughput performance for multi-floor study cases**

These studies clearly reflect the fact that LTE-U suffers negligible interference in Wi-Fi deployment scenarios but the reverse is not true. Wi-Fi suffers huge losses in throughput because its carrier sensing protocol remains on the LISTEN mode a little more than often in presence of LTE-U.

One approach to mitigate interference can be Enhanced Inter-Cell Interference Coordination (eICIC) techniques [6]. The overall objective of the eICIC is to mute certain subframes of one layer of cells in order to reduce the interference to the other layer in heterogeneous network interference scenarios. These muted subframes are called in the LTE standard as Almost Blank Subframes (ABS). ABS together with other strategies to control the mutual interference (e.g., SINR-based power control from LTE side, adaptive channel sensing thresholds from Wi-Fi side and spatial diversity exploitation through multiple antennas techniques) can be combined with some coordination to manage the coexistence of LTE and Wi-Fi systems.

Another way to reduce interference with Wi-Fi is to implement a listen-before talk scheme as shown in Fig. 11. In the listen-before-talk scheme, a device listens to the channel for a period of time and if no ongoing transmission is observed, the device can start its transmission. Prior to downlink data transmission, LTE-U gateway will sense the channel. If available, LTE-U gateway will commence data transmission and inform the users of data transmission. This method is transparent to the UE in the downlink. In the uplink, however, the LTE-U gateway must schedule data transmission *n* subframes ahead of time. In this case, the gateway does not know whether the channel will be free and thus it will be up to the UE to sense the channel before beginning data transmission in the assigned subframe. If the channel is not free, the UE will refrain from transmission. The gateway can then reschedule this transmission using secondary carrier. In the above mechanism, the carrier sensing operation could be different from that of the DCF in Wi-Fi.

The IEEE 802.11 standard also defines the Request-to-Send/Clear-to-Send (RTS/CTS) protocol to provide more reliability to the transmission of large data packets [3]. In the RTS/CTS scheme, the source device listens to the channel for a period of time and if no ongoing transmission is observed, the device sends a RTS message to the destination. The destination responds with CTS and then data transmission can occur. Through the RTS and CTS packets the source and destination devices reserve the channel for the entire duration of the data and acknowledgment packet exchange.

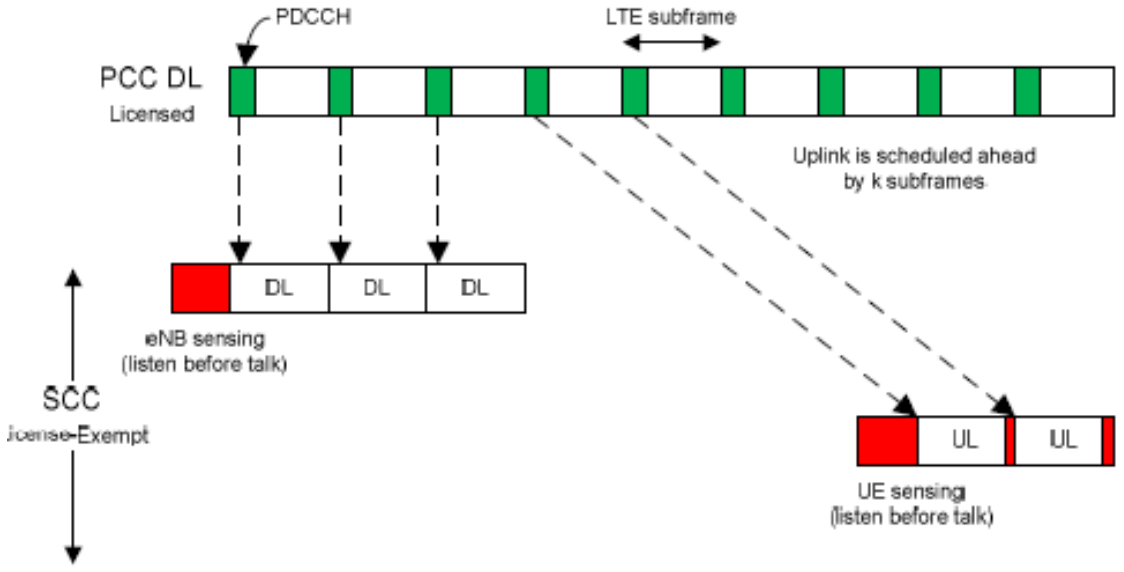

**Fig 11: LTE-U with Listen-before-talk [3]**

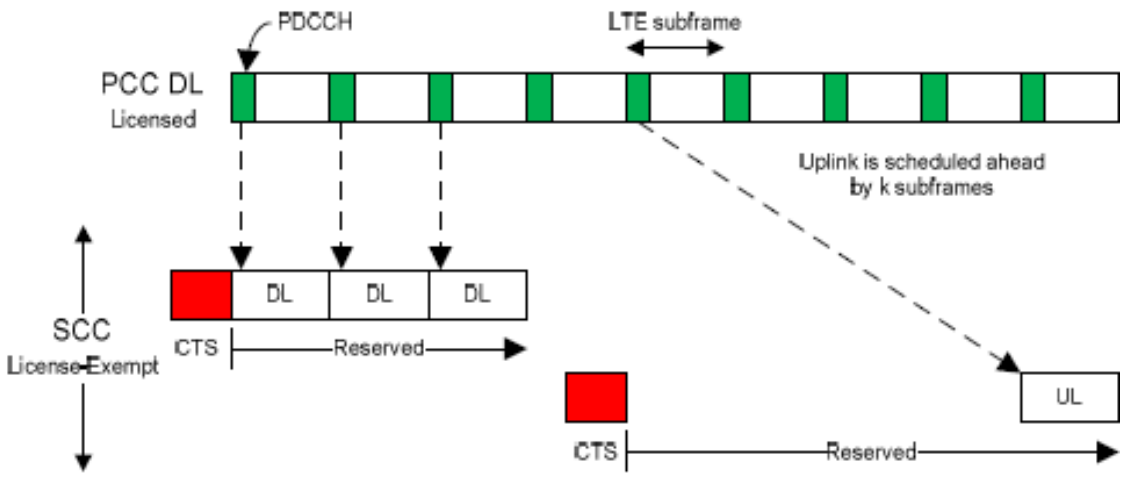

**Fig 12: LTE-U with RTS/CTS mechanism**

Simsek, Bannis [12] also discuss about a possibility of combining small cells with Wi-Fi through cross-system learning framework allowing small cells to optimize their performance where data traffic is offloaded to Wi-Fi and delay sensitive transmissions are retained with the small cell dynamically. Although, the authors claim that the proposed cross-system learning outperforms several benchmark algorithms and traffic steering policies, with gains reaching up to 200% when using a traffic-aware scheduler as compared to the classical proportional fair (PF) scheduler, it will introduce a lot of overhead on our proposed LTE-U Gateway as it is managing several LTE-U micro cells simultaneously.

## VII. AUTHORIZATION ACCESS TO LTE-U

The mechanism to introduce the authorized access is implemented at LTE-U Gateway which takes care of a bunch of LTE-U microcells in vicinity. The whole process can be divided into three steps:

a) *Micro-cell Registration and User Authentication*
In case of a microcell owned by someone other than the operator, for eg. a microcell at home, or a microcell in a café, it has to register at the nearest LTE-U Gateway so that it is taken into consideration while implementing handover algorithm. Moreover, the owner of the microcell has to give the information of individuals who can be given access to the microcell. User can add individuals at a later stage also.

b) *Authorizing Access while initiating handoff*
When a user comes in vicinity of LTE-U microcell, the LTE-U Gateway after checking a bunch of parameters (discussed above) finally checks whether the user is registered for the microcell. If he/she is, then the handoff occurs otherwise user continues to maintain connection with the macrocell eNB.

c) *Authorizing temporary access*
Another user who is not registered for a particular microcell can ask for permission to access the resources of microcell from the admin (user who owns the microcell) and admin can grant temporary access to that user for the duration of the call or data session. Interestingly, the time it takes from user asking for permission and admin granting it won't affect the user's communication since it is already connected to eNB and only initiates a handoff when it gets the access.

## VIII. CONCLUSION AND FUTURE SCOPE

It has been shown that LTE-U is an attractive option to extend the coverage provided by LTE and LTE-A networks to areas with less or no coverage as well as to extend the coverage inside homes and buildings. Although LTE-U itself offers many advantages including higher throughput and spectral efficiency compared to Wi-Fi, its co-existence with Wi-Fi is a matter of concern. But this can be done by introducing proper modifications in LTE-U as explained section VI. We also introduced and simulated a fuzzy logic based algorithm for Handoff decision making in the LTE-U system.

The integration of LTE-U with LTE and LTE-A networks and its co-existence with Wi-Fi is the stepping stone to 5G which will be based on multiple technologies co-existing side-by-side. The fuzzy logic based handover algorithm could be extended to a more intelligent algorithm to select the appropriate Radio access technology (RAT) in a Heterogeneous Networks (HetNets) in 5G [1]. The decision would then have to be based on number of other factors.

## REFERENCES

[1] Kaloxylos, A.; Barmpounakis, S.; Spapis, P.; Alonistioti, N., "An efficient RAT selection mechanism for 5G cellular networks," Wireless Communications and Mobile Computing Conference (IWCMC), 2014 International , vol., no., pp.942,947, 4-8 Aug. 2014

[2] Aziz, A.; Rizvi, S.; Saad, N.M., "Fuzzy logic based vertical handover algorithm between LTE and WLAN," Intelligent and Advanced Systems (ICIAS), 2010 International Conference on , vol., no., pp.1,4, 15-17 June 2010

[3] U-LTE: Unlicensed Spectrum Utilization of LTE - Huawei [Online; accessed 10/18/2014]. [Online]. Available http://www.huawei.com/ilink/en/download/HW_327803

[4] Extending LTE Advanced to unlicensed spectrum Qualcomm [Online; accessed 10/28/2014]. [Online]. Available http://goo.gl/oedLtU

[5] Rahman, M.I.; Behravan, A.; Koorapaty, H.; Sachs, J.; Balachandran, K., "License-exempt LTE systems for secondary spectrum usage: Scenarios and first assessment," New Frontiers in Dynamic Spectrum Access Networks (DySPAN), 2011 IEEE Symposium on , vol., no., pp.349,358, 3-6 May 2011

[6] Ratasuk, R.; Uusitalo, M.A.; Mangalvedhe, N.; Sorri, A.; Iraji, S.; Wijting, C.; Ghosh, A., "License-exempt LTE deployment in heterogeneous network," Wireless Communication Systems (ISWCS), 2012 International Symposium on , vol., no., pp.246,250, 28-31 Aug. 2012

[7] Yusof, A.L.; Zainali, M.A.; Mohd Nasir, M.T.; Ya'acob, N., "Handover adaptation for load balancing scheme in femtocell Long Term Evolution (LTE) network," Control and System Graduate Research Colloquium (ICSGRC), 2014 IEEE 5th , vol., no., pp.242,246, 11-12 Aug. 2014

[8] Chowdhury, M.Z.; Won Ryu; Eunjun Rhee; Yeong Min Jang, "Handover between macrocell and femtocell for UMTS based networks," Advanced Communication Technology, 2009. ICACT 2009. 11th International Conference on , vol.01, no., pp.237,241, 15-18 Feb. 2009

[9] Hasan, M.K.; Saeed, R.A.; Abdalla, A.; Islam, S.; Mahmoud, O.; Khalifah, O.; Hameed, S.A.; Ismail, A.F., "An investigation of femtocell network synchronization," Open Systems (ICOS), 2011 IEEE Conference on , vol., no., pp.196,201, 25-28 Sept. 2011

[10] Habeeb, A.A.; Qadeer, M.A., "Interference evaluation and MS controlled handoff technique for femtocell," Internet, 2009. AH-ICI 2009. First Asian Himalayas International Conference on , vol., no., pp.1,5, 3-5 Nov. 2009.

[11] Cavalcante, A.M.; Almeida, E.; Vieira, R.D.; Chaves, F.; Paiva, R.C.D.; Abinader, F.; Choudhury, S.; Tuomaala, E.; Doppler, K., "Performance Evaluation of LTE and Wi-Fi Coexistence in Unlicensed Bands," Vehicular Technology Conference (VTC Spring), 2013 IEEE 77th , vol., no., pp.1,6, 2-5 June 2013

[12] Simsek, M.; Bennis, M.; Debbah, M.; Czylwik, A., "Rethinking offload: How to intelligently combine WiFi and small cells?," Communications (ICC), 2013 IEEE International Conference on , vol., no., pp.5204,5208, 9-13 June 2013